\documentclass[final,5p,times,twocolumn]{elsarticle}

\usepackage{amssymb}

\usepackage{siunitx}
\DeclareSIUnit[number-unit-product = {}] {\nucleon}{u}
\DeclareSIUnit[number-unit-product = {}] {\c}{c}
\usepackage{xspace} 
\usepackage{xcolor}

\usepackage[pagewise]{lineno}

\biboptions{compress}

\usepackage{calrsfs}
\DeclareMathAlphabet{\pazocal}{OMS}{zplm}{m}{n}

\usepackage{hyperref}

\usepackage{mathtools}
\DeclarePairedDelimiter\abs{\lvert}{\rvert}%
\makeatletter
\let\oldabs\abs
\def\abs{\@ifstar{\oldabs}{\oldabs*}}

\newcommand{\rev}[1]{#1}
\newcommand{\code}[1]{\texttt{#1}}
\def\amev#1{\SI[per-mode=symbol]{#1}{\mega\electronvolt\per\nucleon}}

\journal{Physica Medica}

\begin{document}

\begin{frontmatter}

\title{Preliminary results in using Deep Learning to emulate BLOB, a nuclear interaction model}

\author[1,2] {A.~Ciardiello}
\author[3]{M.~Asai}
\author[4]{B.~Caccia}
\author[5]{G.~A.~P.~Cirrone}
\author[5]{M.~Colonna}
\author[3]{A.~Dotti}
\author[1,2]{R.~Faccini}
\author[1,2] {\\S.~Giagu}
\author[1,2] {A.~Messina}
\author[6]{P.~Napolitani}
\author[5]{L.~Pandola}
\author[3]{D.~H.~Wright}
\author[1,2] {C.~Mancini-Terracciano}
\ead{carlo.mancini.terracciano@roma1.infn.it}

\address[1]{Dip. Fisica, Sapienza Univ. di Roma, Rome, Italy}
\address[2]{INFN Sezione di Roma, Rome, Italy}
\address[3]{SLAC National Accelerator Laboratory, Menlo Park, United States}
\address[4]{National Center for Radiation Protection and Computational Physics, \\ Istituto Superiore di Sanità, Italy}
\address[5]{INFN, Laboratori Nazionali del Sud, Catania, Italy}
\address[6]{Universit\'e Paris-Saclay, CNRS/IN2P3, IJCLab, 91405 Orsay, France}
\begin{abstract}

\textit{Purpose:} 
A reliable model to simulate nuclear interactions is fundamental for Ion-therapy. We already showed how BLOB (``Boltzmann-Langevin One Body''), a model developed to simulate heavy ion interactions up to few hundreds of \amev{}, could simulate also $^{12}$C reactions in the same energy domain.
However, its computation time is too long
 for any 
 medical application. For this reason we present the possibility of emulating it with a Deep Learning algorithm.
 
%
\textit{Methods:} The BLOB final state is a Probability Density Function (PDF) of finding a nucleon in a position of the phase space. We discretised this PDF and trained a Variational Auto-Encoder (VAE) to reproduce such a discrete PDF. As a proof of concept, we developed and trained a VAE to emulate BLOB in simulating the interactions of $^{12}$C with $^{12}$C at \amev{62}. To have more control on the generation, we forced the VAE latent space to be organised with respect to the impact parameter ($b$) training a classifier of $b$ jointly with the VAE.

\textit{Results:} The distributions obtained from the VAE are similar to the input ones and the computation time needed to use the VAE as a generator is negligible. 

\textit{Conclusions:}  We show that it is possible to use a Deep Learning approach to emulate a model developed to simulate nuclear reactions in the energy range of interest for Ion-therapy. We foresee the implementation of the generation part in C++ and to interface it with the most used Monte Carlo toolkit: Geant4.

\end{abstract}


\end{frontmatter}


\section{Introduction}
Ion-therapy is a technique that aims at treating tumour deeply located in the patient body exploiting the ions characteristic dose deposition shape, with the peak at the end of their range, the so-called Bragg peak. It is performed mainly with protons but also with heavier ions, like $^{12}$C.

\rev{Having reliable nuclear fragmentation models in MC simulation toolkits is of utmost 
importance for Ion-therapy~\cite{Amaldi:2005dm} 
as they
 are considered the gold standard for dosimetric calculations~\cite{Rogers:2006ip};
 they are used to generate the input parameters of the treatment planning algorithms~\cite{Parodi:2012fz} and  to validate the dose calculation of such algorithms, especially in cases with large tissue 
      heterogeneities~\cite{Molinelli:2013fg}.
Finally, a large effort is ongoing to develop detectors to measure the radiation emitted during the treatment to allow a non-invasive on-line monitoring of the treatment itself, see for instance~\cite{Mattei:2016vc, toni, michela, giacomo}, and MC calculations are needed to infer the delivered 
dose from the observed spectra~\cite{Battistoni:2008im, Kraan:2015bq}}

Geant4~\cite{Geant4} is one of the most widely used MC toolkits, also for medical 
applications. It is written in C++ and takes advantage of its object-oriented coding paradigm. Geant4 also exploits the multithread capabilities of C++11, allowing an efficient use of modern CPUs.
It is developed by a large international 
collaboration and distributed with an open source licence, which allows also to develop wrappers around it. 
In the last years many programs dedicated to MC medical simulations have been developed wrapping Geant4, and then using its Physics models, such as
GATE~\cite{gate}, GAMOS~\cite{gamos}, and TOPAS~\cite{topas}. The latter in particular is dedicated to Ion-therapy simulations.

Finally, Geant4 has the capability to simulate the body of a specific patient
importing his  Computed 
Tomography (CT) scan in DICOM format~\cite{g4-medical}. 

\rev{Many critical aspects must be taken into account when modelling the therapeutic ion beams, e.g. although elastic and multiple Coulomb scattering events are negligible for charged particles,  they contribute to dosimetric uncertainty especially in oncological applications, because they cause beam widening~\cite{10.3389/fonc.2015.00150}. However, one of the main uncertainties comes from nuclear interaction models.
Moreover, while there are several  models for electromagnetic interactions in the Geant4 package \cite{john,APOSTOLAKIS2009859,2005NIMPB.227..420M},
 there is no dedicated model to describe inelastic nuclear reactions below \amev{100}.
}
It is possible to use models developed to simulate nuclear reactions at higher energies but
\rev{recent literature has shown their limitations in reproducing the measured 
secondary yields in terms of production rates and angular distributions, see for instance~\cite{Braunn:2013dmt,DeNapoli:2012bs,doudet,Bohlen:2010bg}. 
The most recent benchmark is the one developed by the Geant4 Medical Physics Benchmarking Group~\cite{g4-med-bench}. Such a benchmark does not include the ``Quantum Molecular Dynamics'' (QMD)~\cite{Koi10} model, as it has to be forced to work below \amev{100}, but a recent benchmark of it, with the same dataset, can be found in~\cite{ManciniTerracciano:2018bi}.}

To fill this gap we interfaced 
BLOB (``Boltzmann-Langevin One Body'')~\cite{blob} with Geant4. BLOB is a model designed and developed to simulate heavy
ion reactions from the Coulomb barrier up to few
hundreds of \amev{} and we showed that it can be used also to simulate lighter ion, such as $^{12}$C, interactions~\cite{G4BLOB}. 
However, the BLOB computation time is of the order of tens of minutes per interaction, too large for any practical application. To overcome this obstacle we are exploring  the possibility of substituting BLOB with a Deep Learning algorithm, namely a Variational Auto-Encoder (VAE). 

\rev{An alternative could be porting BLOB to GPU, profiting from their single instruction on multiple data approach which would fit very well with the test particles approach used in BLOB. However, porting the code to GPU means to completely redesign and rewrite the code, to fully exploit the GPU capabilities, therefore the developing time required is much larger than developing a Deep Learning algorithm with the existing libraries, like TensorFlow~\cite{tensorflow} (especially with Keras~\cite{keras}) and PyTorch~\cite{pytorch}, to emulate the model. Finally, by porting the code to GPU we do not expect a speed-up greater than 3 orders of magnitudes,  while the Deep Learning approach could result in an acceleration of 9 orders of magnitude~\cite{billion}, so fast to be used also in the fast Monte Carlo codes, developed on purpose for Ion-therapy, such as Fred~\cite{fred}.}

Several Deep Learning algorithms are powerful probabilistic generative models that, after being trained on real examples, are able to produce realistic synthetic samples, and VAE are a class of them~\cite{DL-book}.

 \begin{figure}[!bht]
\centering
    \includegraphics [width=.9\columnwidth]{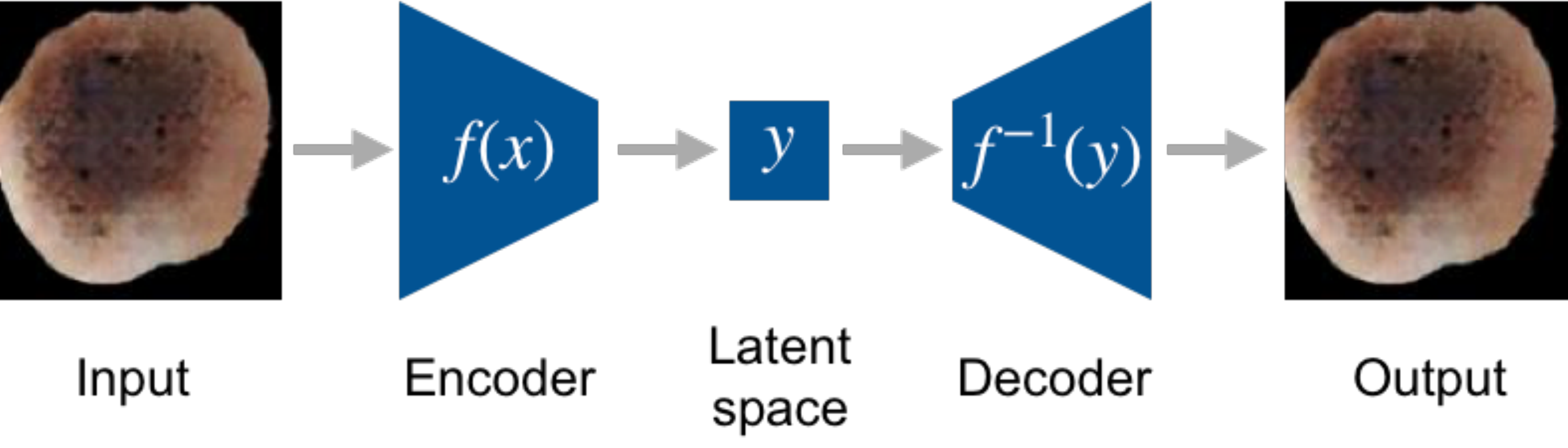}    
\caption{A schematic representation of an AE. An input image, 
\rev{here a melanoma produced with a generative Deep Learning algorithm~\cite{DBLP:journals/corr/abs-1804-03700}, 
}
is input of the encoder, which finds a representation of the input in the latent space. Subsequently, a decoder samples points in the latent space to produce and output image. Usually the encoder and the decoder are symmetric. The AE is trained to produce an output as similar as possible to the input, therefore it can be seen as an identity function.}
 \label{fig:AE}
 \end{figure}

VAE are an evolution of Auto-Encoders (AE) which, in turn, are a class of Deep Learning algorithms trained to reproduce their input as closely as possible. As shown in \autoref{fig:AE}, 
\rev{AE are made by an encoder and a decoder, the encoder 
assignes
a point in the latent space for each input;
 the decoder,
starting from each point in the latent space,
  produces an output with the same dimensionality of the encoder input. 
The encoder and the decoder are jointly trained so that the output, generated from the decoder starting from the latent space point where the encoder maps a given input, is as similar as possible to the input itself.
To make a similarity, the encoder can be seen as 
 an injective function ($y=f(x);\; x \in X ,\, y \in Y$ ), as it maps each distinct point of its domain $X$ in a distinct point of the codomain $Y$.
Once trained, the decoder is like the inverse of the previous function $x=f^{-1}(y)$. Indeed, giving  in input to $f^{-1}$ the point  $y^{\prime}$ in the codomain where $f$ maps an input point $x^{\prime}$ of the domain, 
it gives back $x^{\prime}$ itself as output, i.e. $ x^{\prime} = f^{-1}(y^{\prime})$. Therefore, the AE is trained to be an identity function as it is like the subsequent application of an injective function and its inverse, i.e. $ x^{\prime}=f^{-1}\left(f(x^{\prime})\right)$.}

Being the latent space dimensionality smaller than the input one, the AE is forced to ``learn'' features from the input. VAEs differs from AEs because each input is represented by a distribution probability in the latent space, and not  just by a point, in this way similar inputs are encoded in points close to each other. When used as generative methods this regularisation helps the network to decode plausible outputs from every point of the latent space. In this way, once trained, the VAE -in particular the decoder- can be used to emulate BLOB sampling from the latent space.

\rev{
The main alternative generative model to a VAE is a Generative Adversarial Network (GAN)~\cite{GAN}. It consists of two separate models that are trained in an ``adversarial'' fashion: a generative model produces synthetic data and a discriminative model tries to identify the synthetic data from the real training data.
The two models are trained simultaneously. 
For image generation GAN are often preferred to VAE because the intrinsic probabilistic nature of VAE output tends to generate blurry samples~\cite{revVAE}. However, in our case, even if we chose a neural network architecture derived from image generation (to be precise from video generation), we work, as we will describe lather on, with Probability Density Functions (PDFs), therefore this is not a weakness. Moreover, training a VAE is simpler and requires less data than a GAN, it has a clear probabilistic formulation, and it is easier to introduce priors~\cite{revVAE}. 
 In addiction, a VAE architecture provides an explicit latent representation that can be interpreted as a  dimensional reduction of the input space and one of our  interests  is to explore how the BLOB outputs are encoded in the latent space. This will allow to perform a sampling strategy from the latent space, which in the near future will allow to control the sampling of the impact parameter and in the next steps, to control also the energy of the projectile or the mass number and the charge of the projectile and the target.
}

\section{Material and Methods}
\subsection{Nuclear Interaction Models}
In Geant4, as in many other codes, nuclear reactions are simulated in two steps. The first one describes the dynamic part of the reaction, from the moment in which the projectile and the target collide until the ``thermalisation'' is reached, i.e. when the excitation energy of the large fragments is balanced among all the nucleons composing them. It includes the
pre-equilibrium emissions of small fragments.
The de-excitation phase takes place at this point with the decays of the excited large fragments.

BLOB handles the first part of the nuclear reactions. 
It is based on a semi-classical one-body approach to solve the Boltzmann-Langevin equation. It
describes the time evolution of the nucleon phase space density distribution in a semi-classical way, i.e. taking into 
account the Pauli principle. 
The corresponding transport equation is solved numerically, sampling the density distribution in phase 
space with test particles, whose evolution is done under 
 the action of an effective mean-field nuclear potential.
 Fluctuations in the dynamics are introduced as collisions between the test particles.
 
 \label{par:blobfinalstate}
 The BLOB final state is a Probability Density Function (PDF) of finding a nucleon in a position of the phase space built as a list of the positions of the test particles at the end of the reaction. 
A ``liquid''  and a ``gas'' phase are defined
by applying a clustering
procedure to this PDF.
Each liquid phase neighbourhoods stands for a large fragment.
The rest of the test particles, the gas phase, represents the nucleons emitted in the first part of the reaction.

\begin{figure}[!bht]
\centering
\includegraphics [width=.45\columnwidth]{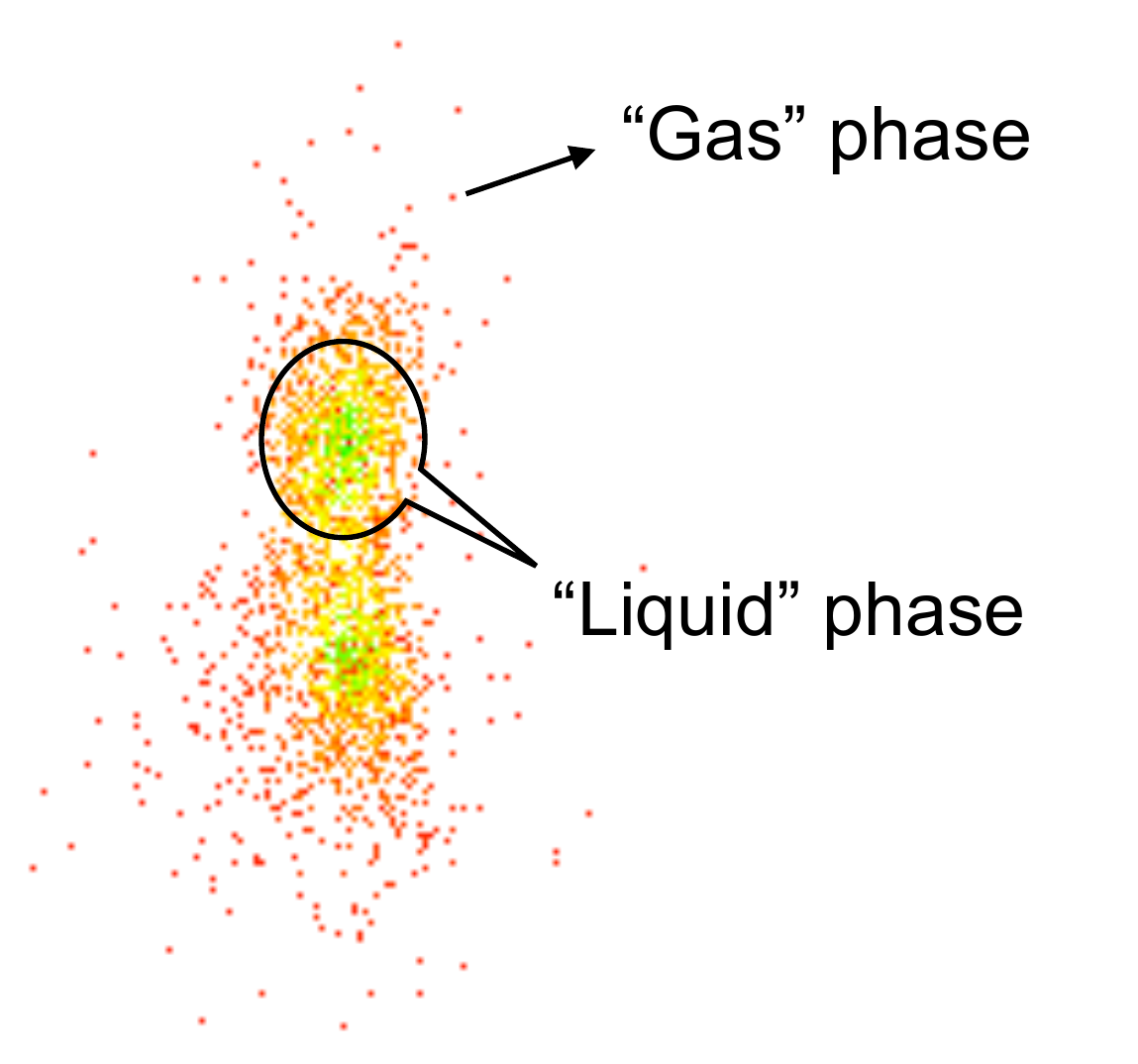}
\caption{\rev{A pictorial representation of a BLOB output, two ``liquid phase'' clusters of test particles are visible in green and many ``gas'' particles emitted are in red. The clustering procedure is described in detail in~\cite{G4BLOB}.
}
}
\label{fig:blobout}
\end{figure}

We already showed in precedent works~\cite{Napolitani:2018ofs,G4BLOB} the BLOB potentialities in describing $^{12}$C fragmentation, 
comparing its predicted yields with experimental data using the SIMON code~\cite{simon} for the 
de-excitation phase  and the Geant4  de-excitation model 
\code{G4ExcitationHandler}~\cite{G4-deexcitation}.

\rev{The BLOB computation time for the interaction of two light (like $^{12}$C) ions at the energy of interest for Ion-therapy  takes order of tens of minutes because, as already mentioned, BLOB relies on the test particles method and, to improve the simulation and the effects of the mean field in particular, we used 500 test particles per nucleon. Therefore the code has to track order of $10^{4}$ test particles to simulate the interaction of two $^{12}$C ions. In addition, in BLOB the potential is self consistent, i.e. it depends on the test particles positions, therefore it has to be computed at each step and its computation time increases linearly with the number of test particles. Moreover, the Pauli blocking term in the collision integral depends on the positions of the couples of test particles and, therefore, it has a quadratic dependence on the number fo test particles.}

  In this work, we present the preliminary results obtained emulating the BLOB model with a Deep Learning algorithm, foreseeing the interface of such algorithm with the Geant4 toolkit.

\subsection{Dimensionality reduction}
\label{sec:dimred}
\begin{figure}[!bht]
\centering
\includegraphics [width=.9\columnwidth]{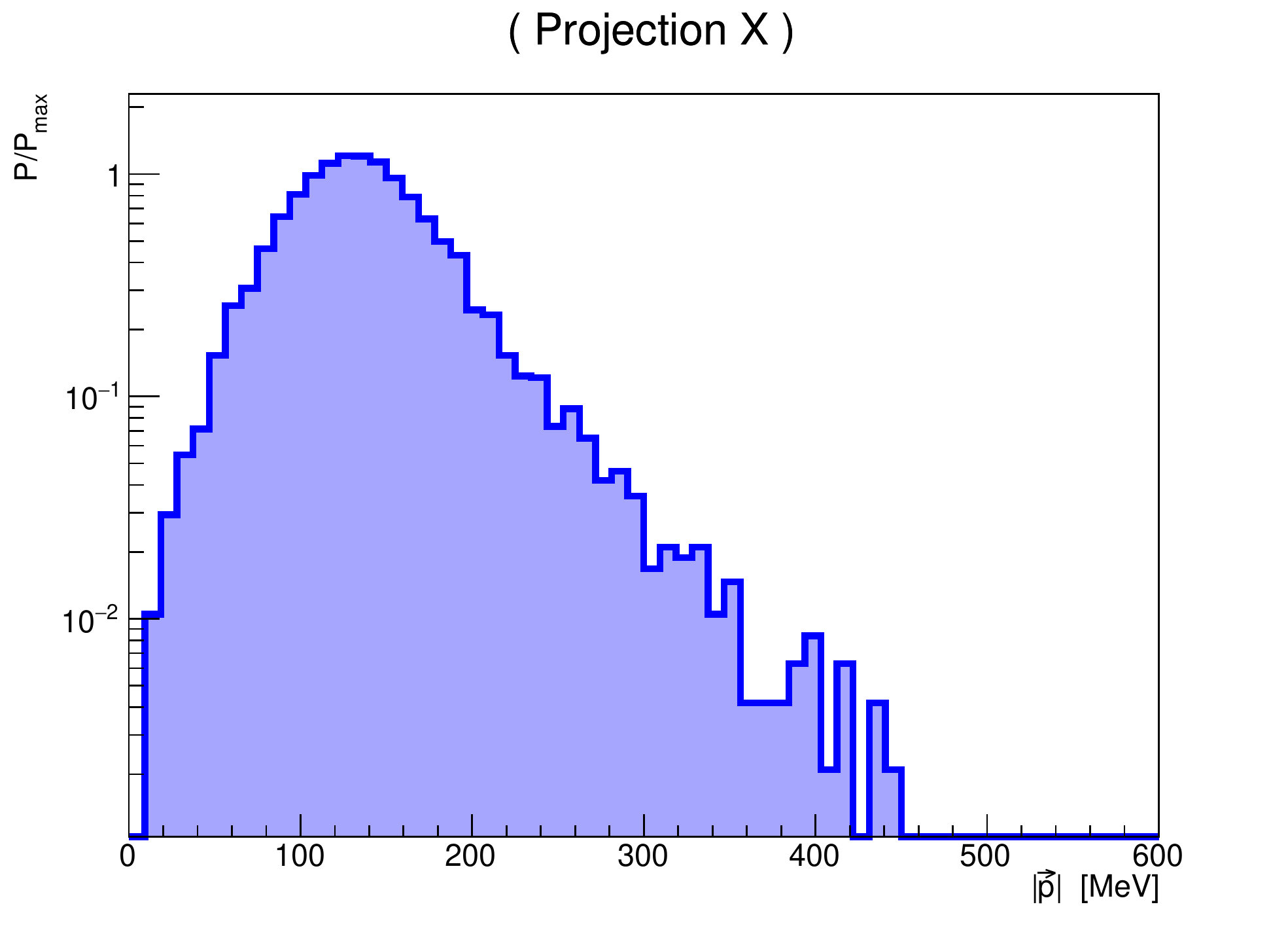}
\includegraphics [width=.9\columnwidth]{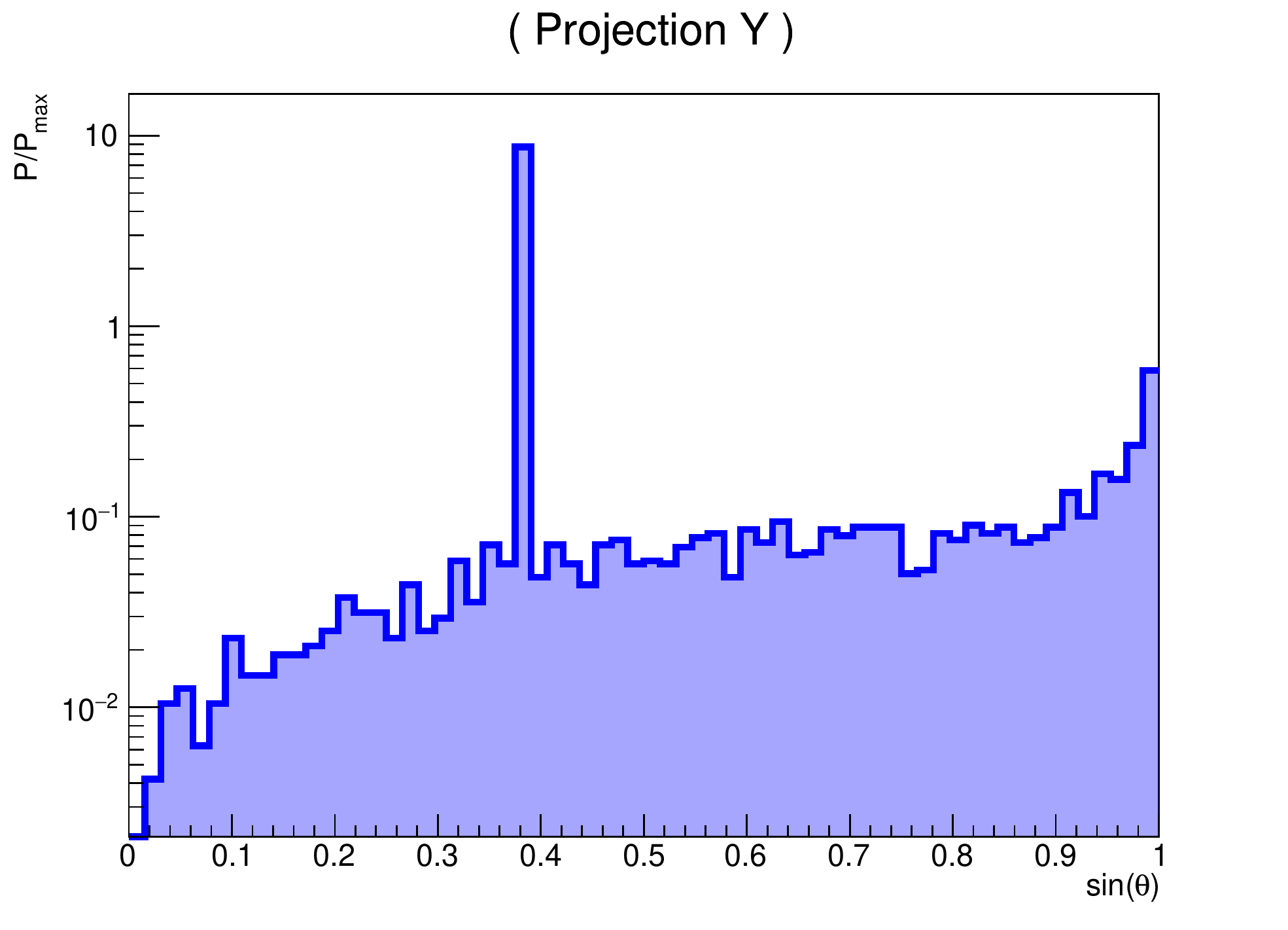}
\includegraphics [width=.9\columnwidth]{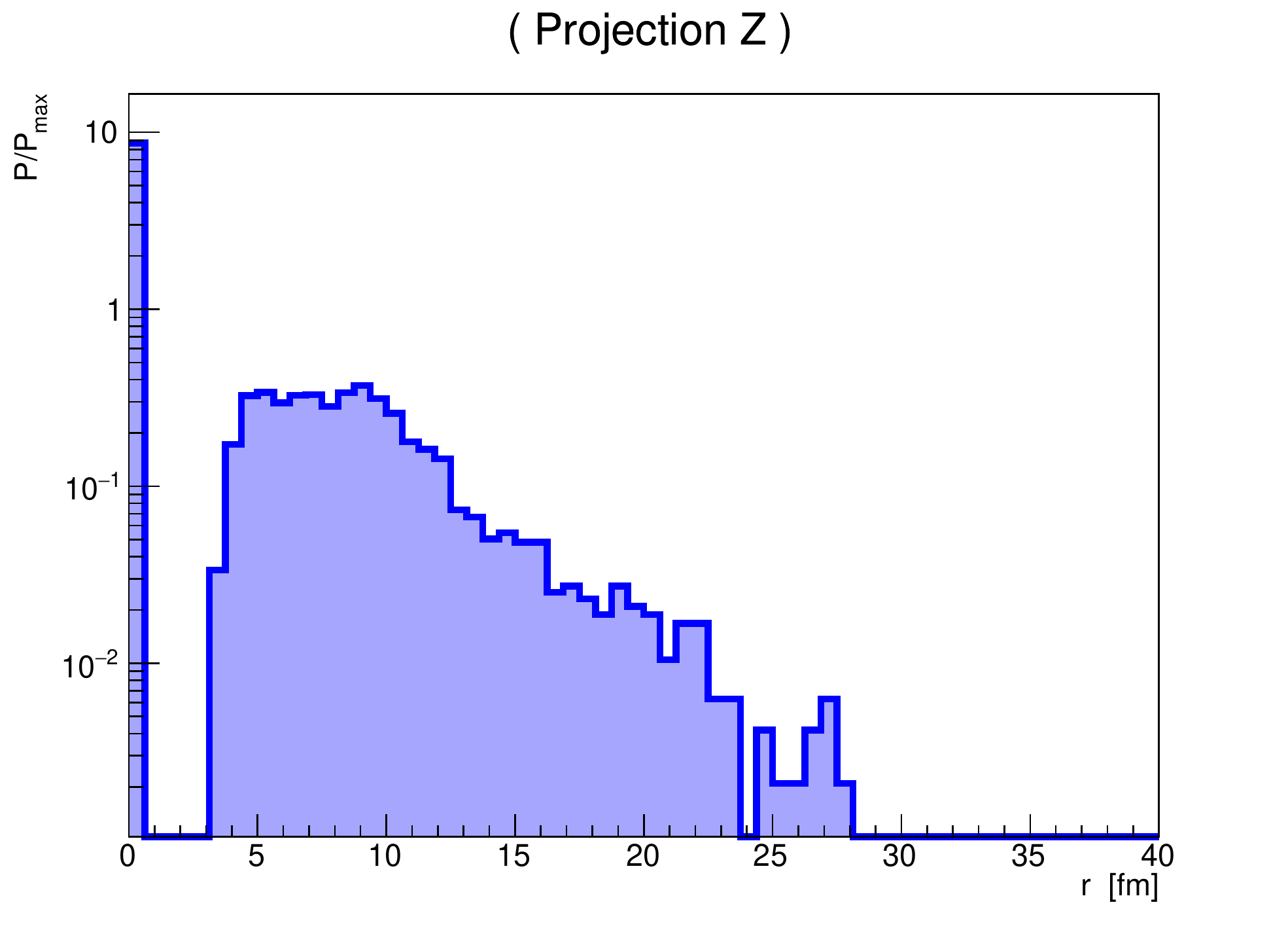}
\caption{Example of one the reduced dimensionality PDF obtained from the BLOB output. The three panels show the projections on the three axis. 
\rev{
As described in the text, the 3 dimensional PDF output of the dimensionality reduction is normalised to its maximum, therefore on the ordinate axes these three plots have a probability ($P$) divided by the maximum probability of the PDF itself ($P_{max}$). The maximum of  the three distributions shown is not 1 because they are projections on the axis (the 3 dimensional PDF has maximum 1).
The peaks at $\sin\left(\theta_{\hat{z}}\right) \approx 0.4$ and  $r=0$  represent the large fragment. 
On the 
$\lvert\vec{p}\rvert$
projection, the large fragment is the gaussian distribution centred roughly at \SI{150}{\mega\electronvolt}.
}}
\label{fig:in}
\end{figure}

\rev{

Let us remind that the BLOB output is a PDF of finding a nucleon in a position of the phase space, therefore it consists of two, one for protons and one for neutrons, six dimensional, three for the spatial coordinates and three for the momenta, distributions.
Such output can be discretised, binning the
phase space.
In theory a VAE can be trained to reproduce these two discretised PDFs,
however, at the moment convolutional six dimensional layers are not available in the most used libraries to develop Deep Learning algorithm, Keras~\cite{keras} and PyTorch~\cite{pytorch}. To perform this feasibility study, we decided to use the existing convolutional layers in Keras. They can handle at most three distributions with three dimensions, as they are made for color videos. Therefore we reduced the dimensionality of the problem from six to three degrees of freedom.
We will show that the information loss do not invalidate the ability of our method to reproduce the BLOB output.
For the reaction under investigation, $^{12}$C on $^{12}$C at \amev{62}, BLOB predicts typically two, and at most three, large fragments in its final state. To simplify this dimensionality reduction, for the moment, we will focus only on events with two large fragments in the final state dividing the BLOB output in two discretised PDF, one per large fragment. 
 We associate each gas particle to one of the two large fragment, in particular to the one with 
the larger value of $\cos\left(\theta_{frag}\right)$, where $\theta_{frag}$ is the angle formed between the gas particle momentum and the fragment momentum. For each gas particle we retain as information: 
 \begin{enumerate}
\item  the modulus of its momentum $\abs{\vec{p}}$; 
\item $\sin\left(\theta_{\hat{z}}\right)$, where $\theta_{\hat{z}}$ is the angle formed between its momentum and the interaction axis $\hat{z}$;
\item the distance $r$ of each gas particle with respect to the large fragment to which it has been associated.
\end{enumerate}
These three variable are the three dimensions of the two PDFs output of the dimensionality reduction, one per each large fragment in the final state.

As  described in the \autoref{par:blobfinalstate}, the large fragments are the ``liquid'' phase identified by a clustering procedure. Therefore, a large fragment stands for many test particles (the mass number of the fragment $A$ is sampled from the number of test particles composing it~\cite{G4BLOB}) and the clustering procedure assigns a position and a momentum to the large fragment, plus an excitation energy. 
To include in the three dimensional PDFs output of this dimensionality reduction the large fragment information, we set the $\sin\left(\theta_{\hat{z}}\right)$ to the value of the large fragment and to 0 the values of $r$. In order to preserve the information about the momentum modulus of the large fragment and its excitation energy, we sampled each entry of momentum modulus from a normal distribution, centred on the momentum modulus of the large fragment itself and with a variance equal to its excitation energy. \autoref{fig:dimred} shows schematically this dimensionality reduction.

\begin{figure}[!bht]
\centering
\includegraphics [width=.9\columnwidth]{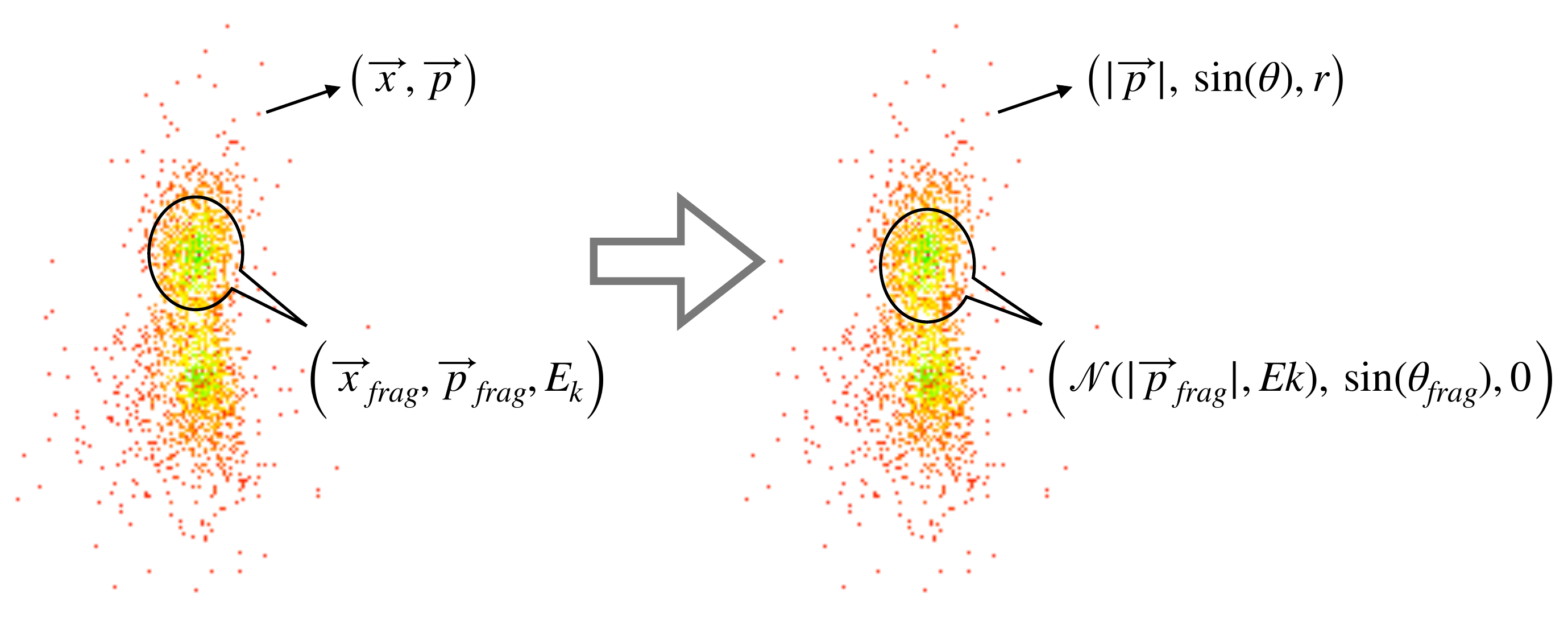}
\caption{\rev{Same BLOB output shown in \autoref{fig:blobout}. The gas test particles can be described by their position and momentum, besides the electrical charge; the large fragments have also an excitation energy. 
The dimensionality reduction describes the positions of all the gas test particles with respect to the closest large fragment in polar coordinates but saving only the distance ($r$), the azimuthal and polar angle are inferred from the direction of flight.
 The $r$ is assumed to be 0 for all the test particles being part of a large fragment itself.
Also the momenta are saved in polar coordinates, saving the modulus of it $\left( \lvert\vec{p}\rvert\right)$, and its angle with the interaction axis, $(\theta)$ (to be precise the $\sin(\theta)$). To keep the momentum and excitation energy of the large fragments we sample the $\lvert\vec{p}\rvert$ 
 from a Gaussian distribution with the fragment momentum as mean and its excitation energy as variance.
}
}
\label{fig:dimred}
\end{figure}

Finally, the PDFs are normalised to the maximum of the two, in order to give in input to the VAE numbers not too small.

\autoref{fig:in} shows the  projection on the three axis of a reduced dimension PDF, i.e. it is one of the two PDFs output of the dimensionality reduction. The peaks at $\sin\left(\theta_{\hat{z}}\right) \approx 0.4$ and  $r=0$ axis represent the large fragment. If we select these entries, we would see on the $\abs{\vec{p}}$ projection a Gaussian distribution centred roughly at 150~MeV, the large fragment momentum.
}

The inverse operation \rev{of the dimensionality reduction} is done starting from the identification of the entries in the discrete reduced PDF relative to the large fragment, which is easily done selecting all the entries of the bin at 0 in the $r$ axis.
Then, the large fragment momentum and excitation energy are computed extracting their average and variance on the $\abs{\vec{p}}$ axis. All the large fragment entries have also only one value in $\sin\left(\theta_{\hat{z}}\right)$. The momentum azimuthal angle is sampled uniformly. 
As we fill the discrete reduced PDF with $\sin\left(\theta_{\hat{z}}\right)$, there is an ambiguity on the direction of the momentum on the $\hat{z}$ axis, but as BLOB simulates the interactions in the center of mass frame, the direction of the first fragment is chosen randomly and the second fragment is assumed to flight in the opposite direction.

The same reconstruction is applied to each gas particle, taking the momentum direction on the $\hat{z}$ axis as the one of the correlated large fragment. 

\rev{Finally, the correct normalisation of the two PDFs, each one associated with a large fragment, is recovered imposing that the sum of the integrals of the two PDFs is 1.}

\begin{figure}[!bht]
\centering
\includegraphics [width=.9\columnwidth]{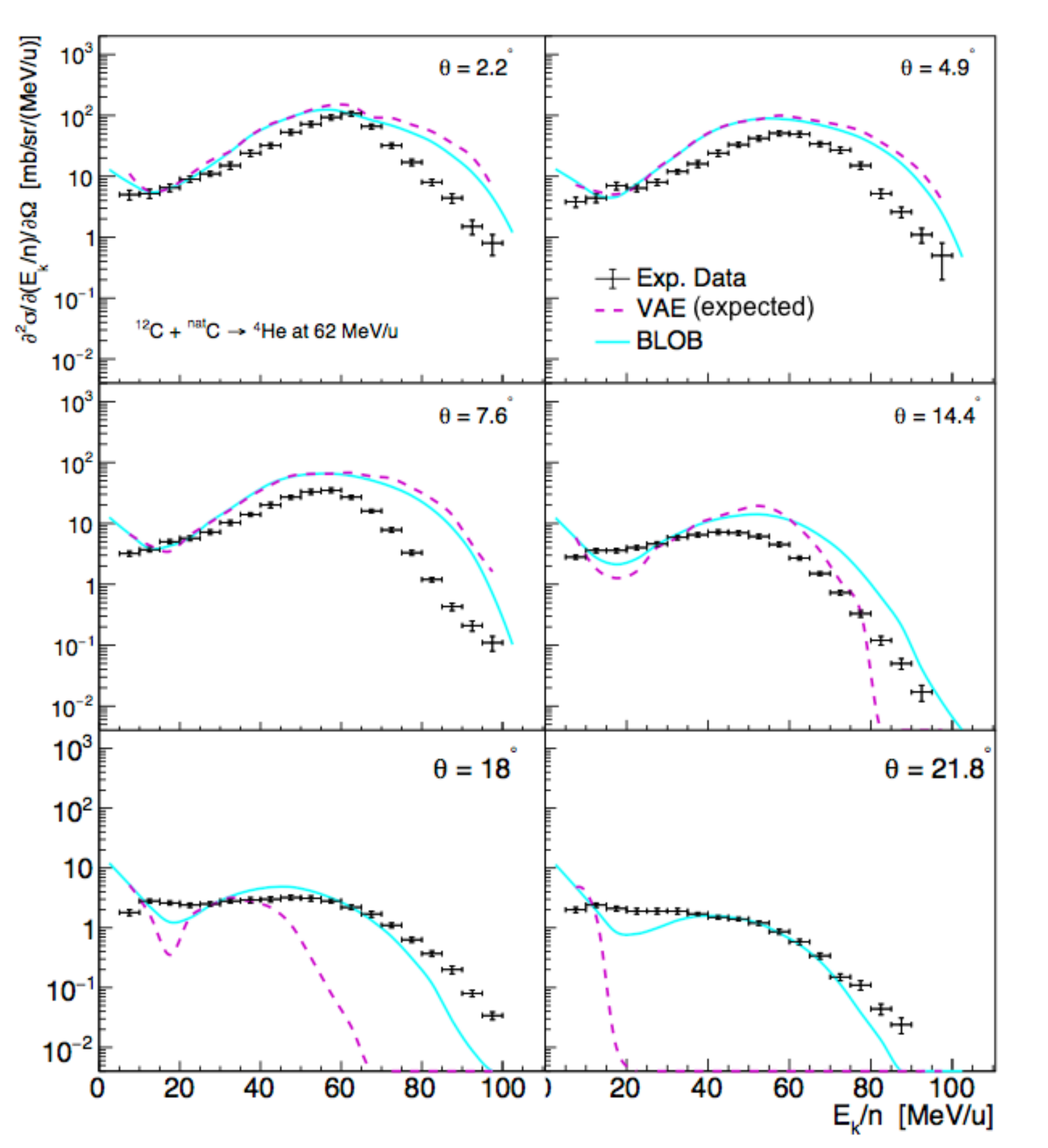}
\caption{Double differential cross sections of alpha particle production in the reaction of $^{12}$C on a thin $^{nat}$C target at \amev{62} as a function of the kinetic energy of the produced fragment for different angles. The experimental data, in black crosses, are from De Napoli et al.~\cite{DeNapoli:2012bs}; the continuous light blue lines show the BLOB predictions and the dashed magenta lines show the calculated values once encoded to reduce the PDF dimensionality and then decoded back. \rev{It is also the expected output of VAE once it will be integrated in Geant4 and interfaced with its nuclear de-excitation phase model.} The differences between the two are due to the approximation made in taking into account events with up to two fragments in the final state. }
\label{fig:testreco}
\end{figure}

\autoref{fig:testreco} shows the results obtained in this dimensionality reduction and its inverse operation. Neglecting events with three large fragments in the BLOB final state produces a lack of high energy ejectiles at large angles, but the distributions up to $7.6^{\circ}$ are well reproduced, enough for the proof-of-concept purpose of this work.

\subsection{Variational Auto-Encoder}
\label{sec:VAE}
As mentioned in the introduction, the VAE is a generative model based on the standard Auto-Encoder that also implements a strong regularisation on the form of the latent variables. This family of models 
are
made by an encoder and a symmetric decoder. 

In our implementation of the VAE, the encoder consists of two  3D convolutional block, each composed by a 3D Convolutional layer, a batch normalisation, a dropout layer with $p=0.2$, and a 3D convolutional downsampling layer with stride 2. Both blocks had a filter size 3x3, and 48 and 24 convolutional kernels respectively.
We used in the first layer an Exponential Linear Unit (ELU) activation function that behaves better with our sparse input and Rectified Linear Units (ReLU) activation function elsewhere because it is one of the less computationally expensive.
The second block is followed by a 3D convolutional layer (24 kernels) and one fully connected layer of width 64 with a hyperbolic tangent as activation function. 
The last layer is  a 2-dimension latent space vector.

At the moment, the latent space is bidimensional and we are obtaining good result, as it will be shown hereinafter, but we plan a dedicated study to verify the best latent space dimension.

The decoder is symmetric with respect to the encoder, using the Keras \code{Conv3DTranspose} layers as upsampling. Also in the decoder, all the layers have a ReLU activation function except the last one, in which we used a Sigmoid, as it is monotonic and has a fixed output range $(0,1)$, perfect to represent probabilities, as in our case.

\subsubsection{Conditioning the latent space}
\label{sec:conditioning}
Taking inspiration from G{\'o}mez-Bombarelli~et~al.~\cite{GomezBombarelli:2018dh}, we trained a classifier for the impact parameter ($b$) jointly with the VAE. Such classifier  tries to predict the impact parameter for each point encoded in the latent space, to force the latent space itself to be organised with respect to $b$. I.e. more likely close points in the latent space encode for events with similar $b$.
To implement this classifier  we used two fully connected layer with 64 neurons.

\rev{The reason of this conditioning is twofold: 
\begin{itemize}
\item Allowing to decide the impact parameter of each event once the VAE will be used to generate events;
\item Showing that it is possible to use this technique to condition the event generation as we plan to use the same technique to train a  VAE with a larger latent space to emulate events at different energies and, in a near future, to allow also to emulate different couples of projectile and target.
\end{itemize}
}

\subsubsection{Loss function, annealing and training}

The loss function for a VAE is usually the sum of a reconstruction term, that express how well the model is able to reconstruct the input data from the compressed latent representation, plus a term that  quantify the distance between the encoded distribution (the posterior) and a normal distribution (the prior). The latter is, as usually done for VAE, a Kullback–Leibler divergence (KL)~\cite{kullback1951}. For the first one, usually called Reconstruction Loss (RC), we chose a Binary Cross-entropy. The two have been linearly combined to be of the same order of magnitude.
The predictor is jointly trained with a mean square error loss function and this loss term is added to the VAE losses. 

To successfully train the model to correctly reproduce the input data, we used an annealing~\cite{bowman-etal-2016-generating} strategy for the KL loss. It consists in multiplying the KL by a weight. At the start of the training this weight is set to zero and after 10 epochs it is linearly increased until it reaches one in 100 epochs.
In this way the model is free to learn to reproduce the input before the conditions on the posterior are applied. 
Intuitively this is equivalent to gradually transform a standard Auto-Encoder to a Variational Auto-Encoder so that when the learned distribution are forced to match the prior distributions the model is already able to encode in the latent space much of the information of the input.

\rev{Our data set consists of 2000  events generated with BLOB and preprocessed with the dimensionality reduction procedure described in \autoref{sec:dimred}. Such events have been generated} with uniform $b$ distribution. \rev{1500 of them have been used} for training and 500 for validation.

We used Adam as optimisation algorithm and, to mitigate the problem due to the high sparseness of the input and accelerate the convergence of back-propagation learning, we initialised randomly the  weights of each layer using the ``Le Cunn Normal'' function~\cite{efficient-backprop}.

\begin{figure}[!bht]
\centering
\includegraphics [width=.9\columnwidth]{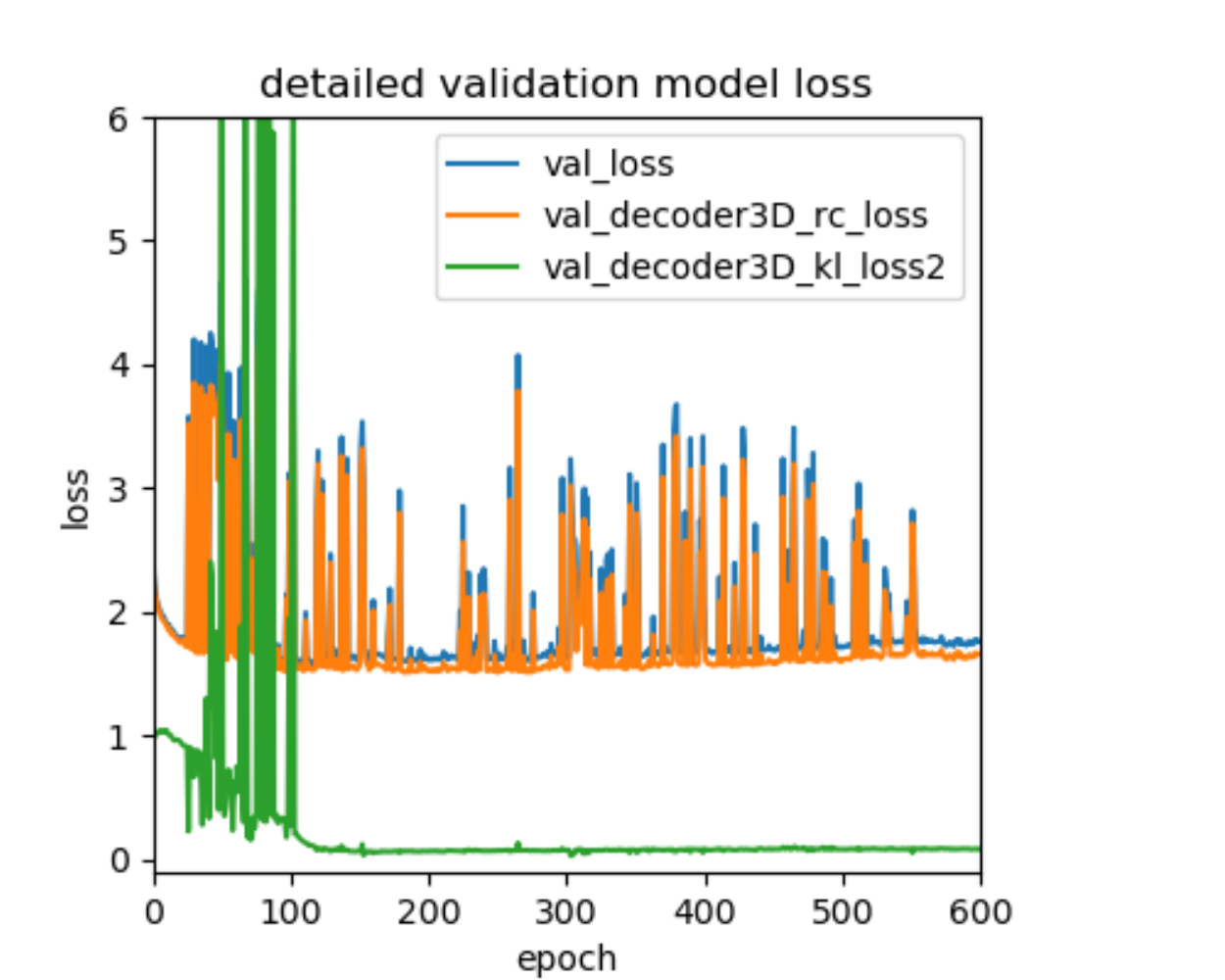}    
\caption{Validation loss as a function of the training epoch, in blue, which is the sum of the RC, in orange,  and the KL, in green, losses. The KL linear decreasing is inversely proportional to the annealing weight, described in the text. 
\rev{The large fluctuations are an effect of the small batch size and the small validation set.  This effect will be mitigated once the training dataset is enlarged. The batch size is limited by the memory of the GPU used for the training.}}
\label{fig:loss}
\end{figure}

\section{Results and Discussion}

\rev{We trained our model for 600 epoch and, as can be seen in \autoref{fig:loss}, we found that it reaches optimal performance within 150-200 epochs. Indeed, \autoref{fig:loss} shows the validation loss, i.e.: the sum of the RC and KL losses calculated on the validation set, as a function of the training epoch and it reaches a plateau roughly after 150 epochs.}

\begin{figure}[!bht]
\centering
\includegraphics [width=.9\columnwidth]{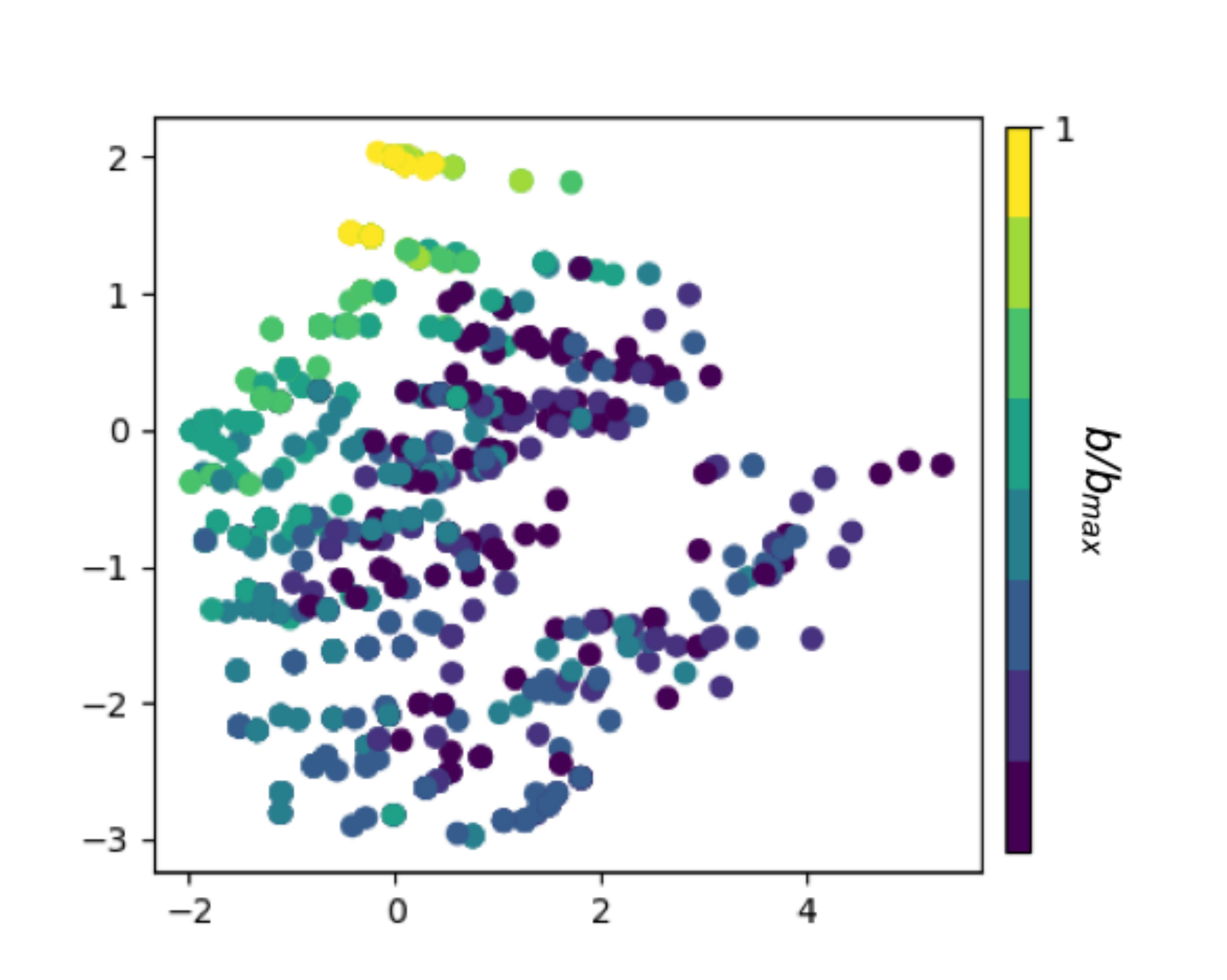}    
\caption{Latent space of the VAE representation. Each point is the encoding point of one training distribution, the color scale represent the impact parameter of the event.}
\label{fig:latent}
\end{figure}

\rev{After such training,} 
we tested that the latent space had been conditioned by the classifier encoding all the training set, the result of this test is shown in \autoref{fig:latent}. Each point is the encoding point of one event and the color scale represents the impact factor ($b$) of each one. The events with large $b$ cluster in the upper part of the plot, for the other events it seems that there is an organisation but we have to improve it enlarging the training set, as the organisation of the latent space with respect to $b$ is fundamental to sample from it when the VAE will be used to generate events. 

\begin{figure}[!bht]
\centering
\includegraphics [width=.9\columnwidth]{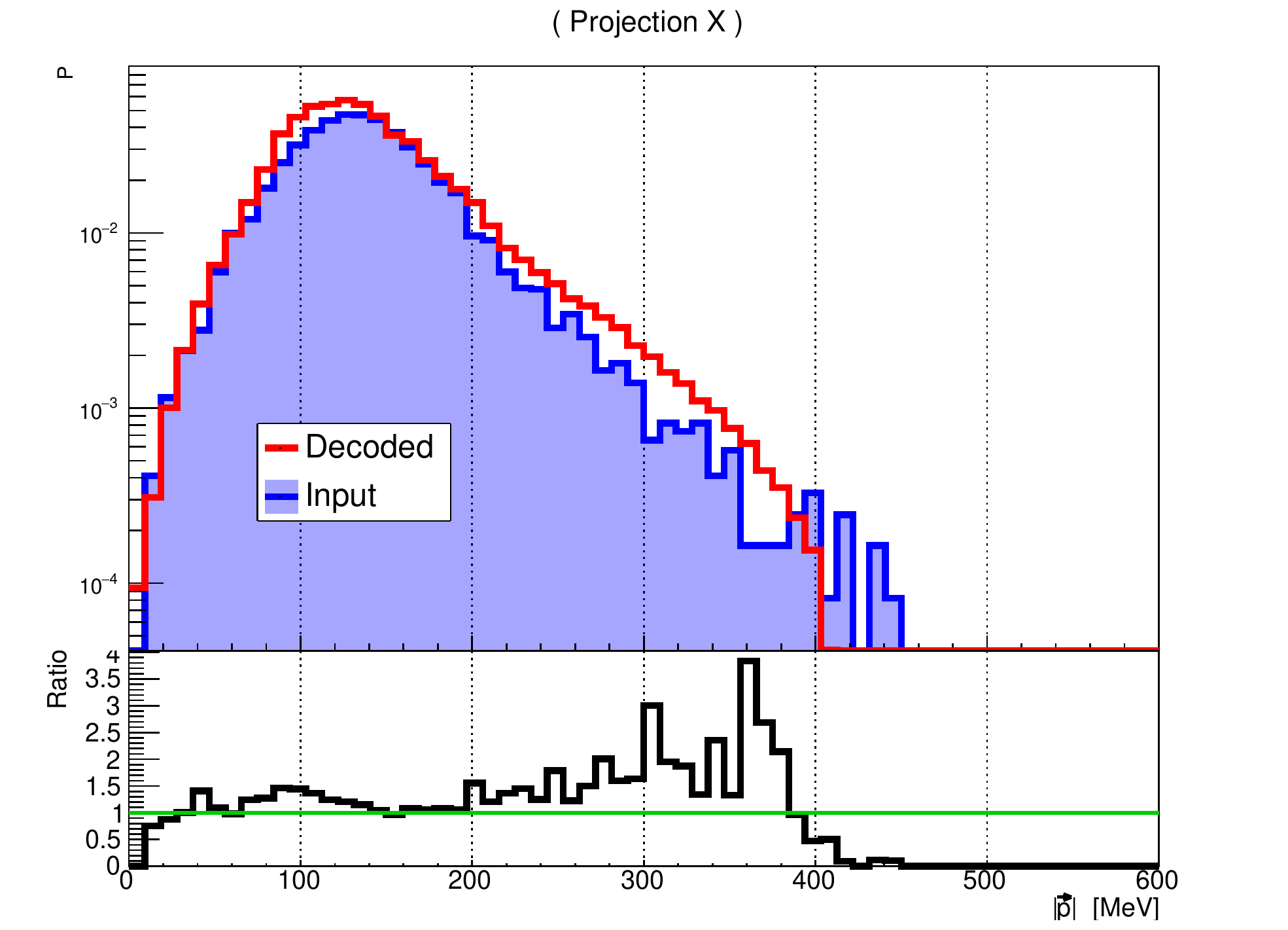}
\includegraphics [width=.9\columnwidth]{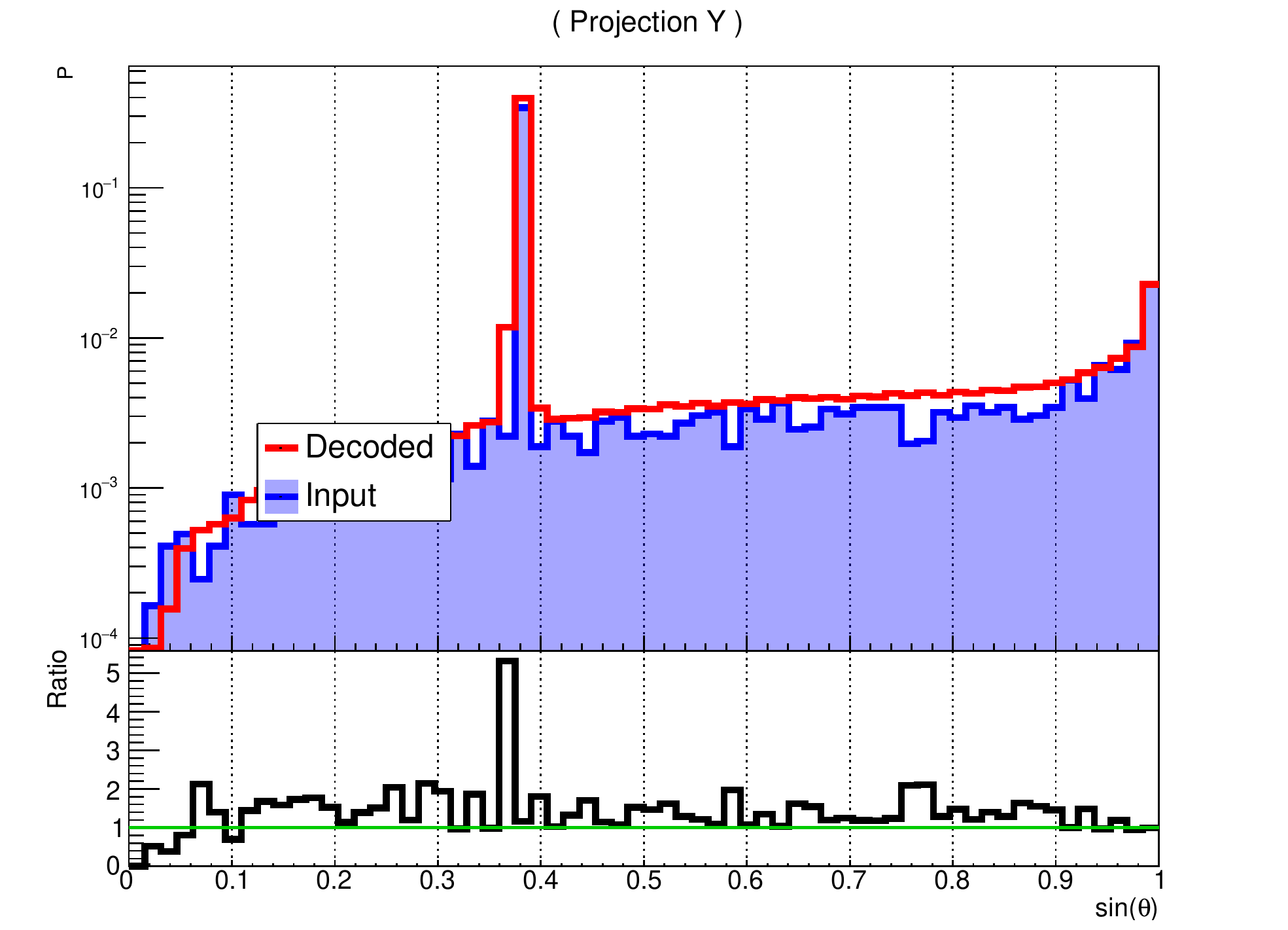}
\includegraphics [width=.9\columnwidth]{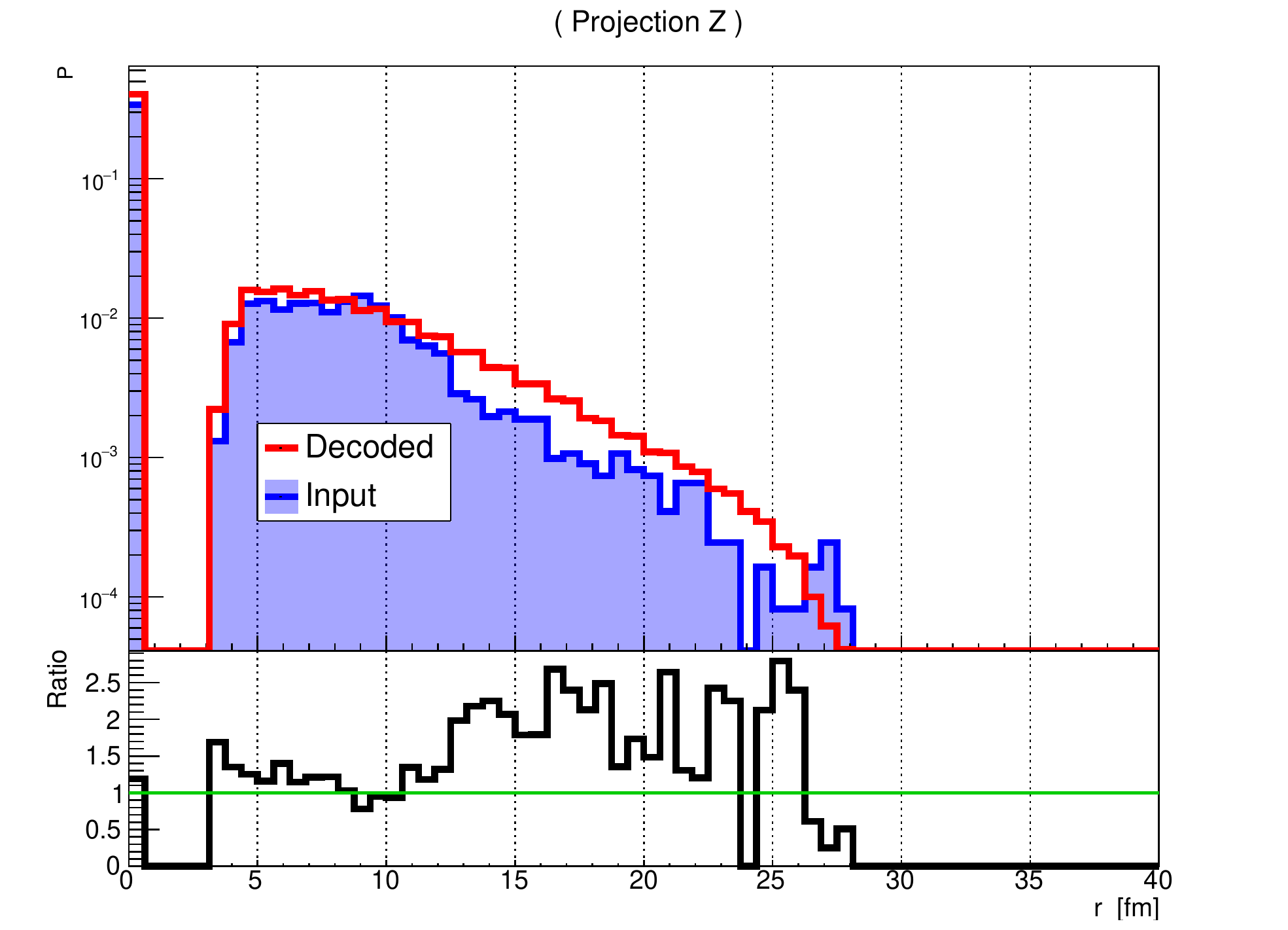}
\caption{\rev{Same as \autoref{fig:in} but on top of the input distribution projections, in blue, the red histograms show the projections of the distribution generated from the decoder part of the VAE. Such an event has been generated starting from a point in the latent space close to the one where the input is encoded. Unlike \autoref{fig:in}, here both the PDFs are probabilities, as they are normalised imposing that the integrals  is 1. The bottom part of each panel is showing the ratio between the generated distribution and the input one. Such ratio deviates significantly from 1 only for the bins with quite low statistics (or close to the peak in the middle panel).}}
\label{fig:out}
\end{figure}

\begin{figure}[!bht]
\centering
\includegraphics [width=.9\columnwidth]{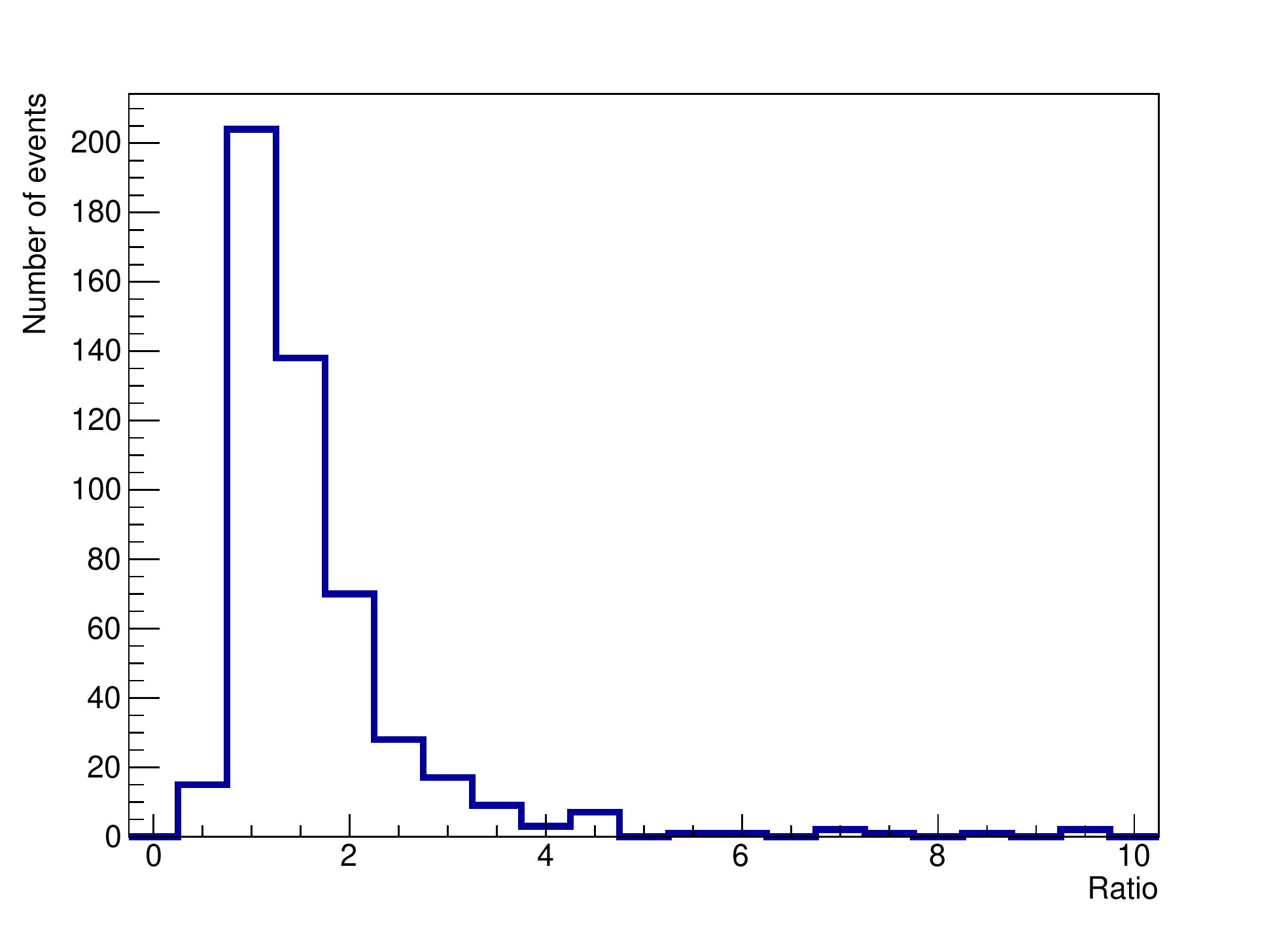}    
\caption{Distribution of the average ratio calculated bin per bin on each output distribution obtained sampling a point in the latent space close to each point where an input from the validation set is encoded. }
\label{fig:ratio}
\end{figure}

To test the VAE generative capabilities, we used it to generate a distribution from a point close to the one where the first event of the validation set is encoded. The generating point in the phase space has been sampled from two normal distributions, one for each axis of the latent space, using as the mean values the coordinates of the encoding point of the testing distribution and as variances 10\% of them. The results are shown in \autoref{fig:out} where it can be seen that the output distribution is very close to the input one. It has to be underlined that we did not use this input distribution for the training phase and so that the VAE never saw this particular event before. We are showing only one of the two color channels because they are pretty similar. \rev{To give an estimation of the similarity of the two distributions, we computed the average ratio on the $\abs{\vec{p}}$ projection, as it is the most significant value from the physical point of view. For the event shown in \autoref{fig:out} the average ratio is 1.06. We performed the same procedure with all the events of the validation set, \autoref{fig:ratio} shows the distribution of the average ratio on all the validation set events. 85\% of the events are in the first four bins, i.e. the average ratio between the VAE output and the closest input is less than 2. To improve this result we need to enlarge the training and validation datasets.}

\section{Conclusions}

We explored the possibility of using a VAE to emulate BLOB, a model made to simulate nuclear interactions below  \amev{100}. The rationale of this proof-of-principle  is that these interactions are of great importance for MC simulations for Ion-therapy and therefore a reliable model is needed but the running time of this kind of models is too large for medical applications. Taking advantage of the fact that the BLOB final state is a PDF and of the recent improvements of the  capabilities of Deep Learning algorithms in generating realistic synthetic samples, we developed and trained a  VAE to emulate BLOB. To use the existing libraries to implement our VAE we reduced the dimensionality of the problem, exploiting the characteristic of $^{12}$C interactions in this energy domain. 

The obtained results are encouraging, however we need to increase the training set to improve the latent space organisation. We aim to sample from the latent space, extracting the impact parameter event by event. We used a limited training set in this work as this is just a proof-of-concept and keeping the training set limited shorten considerably the training time and the computational resources needed.

\rev{As mentioned in \autoref{sec:conditioning}, 
we plan to enlarge the VAE latent space and train the VAE} with different projectile energies and using different ions as projectile and target, adding a classifier {to condition} the latent space for each {degree of freedom introduced. 
In this way a point in the latent space is associated to a desired output properties and
 we will be able to set for each event generated after the training phase, all the characteristics of the projectile and target couple and the interaction energy, as we can do now with the impact parameter.}

Finally, we plan to import the decoder part of the VAE in C++ to interface it directly with Geant4, so that it will be possible to use a state-of-the-art model for low energy nuclear interaction without its computational overhead.


%




\bibliographystyle{elsarticle-num}

\bibliography{VAE}

\end{document}